%% file: jws_ontology_paper.tex
\documentclass[twocolumn]{elsart3p}
\usepackage{epsfig}
\usepackage{amssymb}
\usepackage{url}

\input{inc_review}

\input{inc_symbols}

\usepackage{listings}
\usepackage{color}

\definecolor{lightblue}{rgb}{.3,.5,1}
\definecolor{orange}{rgb}{1,.7,0}
\definecolor{darkorange}{rgb}{1,.4,0}
\definecolor{darkgreen}{rgb}{0,.4,0}
\definecolor{darkblue}{rgb}{0,0,.4}
\definecolor{darkred}{rgb}{.56,0,0}
\definecolor{gray}{rgb}{.2,.2,.2}
\definecolor{lightgray}{gray}{0.7}
\definecolor{shadecolor}{gray}{0.7}

\usepackage[T1]{fontenc}

\lstdefinelanguage{RDF_N3}{
      morekeywords=[1]{@prefix, a },
      morestring=[b]",
      morecomment=[s]{<}{>}, 
      morecomment=[l][\it \color{gray}]{\#}, 
      morecomment=[s][\color{lightgray}]{\^\^}{\ }, 
      otherkeywords={[, ], (, )},%
      sensitive=false%
}[keywords,comments,strings]

\begin{document}
 
\begin{frontmatter}

\journal{Arxiv}
\title{The Research Object Suite of Ontologies: Sharing and Exchanging Research Data and Methods on the Open Web\thanksref{thx1}}
\thanks[thx1]{The Research Object Ontologies have been developed under the aegis of the EU Wf4Ever project (http://www.wf4ever-project.org).}

\author[a]{Khalid Belhajjame}
\author[b]{Jun Zhao}
\author[c]{Daniel Garijo}
\author[d]{Kristina Hettne}
\author[e]{Raul Palma}
\author[c]{Oscar Corcho}
\author[f]{ Jos\'e Manuel G\'omez-P\'erez}
\author[a]{Sean Bechhofer}
\author[b]{Graham Klyne}
\author[a]{Carole Goble}

\address[a]{School of Computer Science, University of Manchester, UK.\\
    first$\_$name.last$\_$name.@cs.man.ac.uk}
\address[b]{Department of Zoology, University of Oxford, UK.\\
	first$\_$name.last$\_$name.@zoo.ox.ac.uk}    
\address[c]{Ontology Engineering Group, Universidad Polit\'{e}cnica de Madrid, Spain.\\ \{dgarijo, ocorcho\}@fi.upm.es}
\address[d]{Leiden University Medical Center, Leiden, The Netherlands. \\
	k.m.hettne@lumc.nl} 
\address[e]{Poznan Supercomputing and Networking Center, Poznan, Poland. \\ rpalma@man.poznan.pl}	
\address[f]{iSOCO, Madrid, Spain. \\ jmgomez@isoco.com}

\begin{abstract}

Research in life sciences is increasingly being conducted in a digital and online environment. In particular, life scientists have been pioneers in embracing new computational tools to conduct their investigations. To support the sharing of digital objects produced during such research investigations, we have witnessed in the last few years the emergence of specialized repositories, e.g., DataVerse and FigShare. Such repositories provide users with the means to share and publish datasets that were used or generated in research investigations. While these repositories have proven their usefulness, interpreting and reusing evidence for most research results is a challenging task. Additional contextual descriptions are needed to understand how those results were generated and/or the circumstances under which they were concluded. Because of this, scientists are calling for models that go beyond the publication of datasets to systematically capture the life cycle of scientific investigations and provide a single entry point to access the information about the hypothesis investigated, the datasets used, the experiments carried out, the results of the experiments, the conclusions that were derived, the people involved in the research, etc. 

In this paper we present the Research Object (RO) suite of ontologies, which provide a structured container to encapsulate research data and methods along with essential metadata descriptions. Research Objects are portable units that enable the sharing, preservation, interpretation and reuse of research investigation results. 
The ontologies we present have been designed in the light of requirements that we gathered from life scientists. They have been built upon existing popular vocabularies to facilitate interoperability. Furthermore, we have developed tools to support the creation and sharing of Research Objects, thereby promoting and facilitating their adoption.

\end{abstract}
 
\end{frontmatter}

\input{introduction-new}
\input{ro-khalid}
\input{related_work}
\input{example-scenario}

\input{ro_ontologies_overview}

\input{ontologies}

\input{tools}

\input{conclusions}

\bibliographystyle{plain}
\bibliography{bib}

\end{document}

%% file: inc_review.tex
\usepackage{color}

\definecolor{darkgreen}{rgb}{0,.5,.2}
\definecolor{gray}{rgb}{.5,.5,.5}
\definecolor{Black}{rgb}{0.000,0.000,0.000} 

{\vspace{5mm}\begin{sloppypar}
  \noindent\hrulefill\, BEGIN \,\hrulefill
  \end{sloppypar}
}%
{\begin{sloppypar}
  \noindent\hrulefill\, END   \,\hrulefill
  \end{sloppypar}\vspace{5mm}
}

\newcounter{todo}

\newcounter{question}

\newcounter{comment}

\newcommand{\change}[1]{{\color{Black}#1}}

%% file: inc_symbols.tex
\newtheorem{definitionTheorem}{Definition}

\newtheorem{exampleTheorem}{Example}



%% file: introduction-new.tex
\section{Introduction}
Research in life sciences is increasingly digital \cite{tompa2005assessing,10.1109/MIC.2012.122}. Life scientists have been pioneers in embracing new computational tools to conduct their investigations. For example, they have adopted scientific workflows as a means for designing and automating the execution of their {\it in silico} experiments \cite{Tiwari2007Workflows}. Life scientists are also one of the main adopters and drivers of semantic web technologies. This is partly witnessed by the numbers of life science datasets that are published on the linked data cloud\footnote{http://linkeddata.org/}. 

Just like in other modern sciences, to report their findings life scientists use scholarly publications as the main trusted means to spread and communicate their results. However, publications are sometimes insufficient to communicate all the scientific knowledge behind the reported results. Scientists often find it difficult (or even impossible) to recover information about the details of an investigation based solely on the published articles \cite{stodden11}. There is a gap, or more specifically knowledge loss, between the environment in which an investigation is carried out, the design and execution of experiments and interpretation of their results, on one hand, and the medium used for disseminating investigation results, i.e., scholarly articles, on the other hand. 

This knowledge gap leads to a set of challenging hurdles. In particular, research findings may not be accurately interpreted without access to comprehensive contextual information like the datasets and experiments used in the investigation. Moreover, scientists are likely to struggle reusing or reproducing existing research results due to the lack of information that connects the concepts and conclusions reported on in the paper to the computational environment used in the investigation and its configuration \cite{DBLP:conf/ic3k/Goble11}.  

To (partly) overcome this problem, we observe that some journals in the area, e.g., the Journal of Biomedical Semantics\footnote{\url{www.jbiomedsem.com}}, require the authors to provides information in dedicated sections in the article specifying the methods they adopted in their research as well as information on where material, such as the software products used in the experiments and the datasets used and generated as a results of the investigation, can be found. Furthermore, life scientists have started adopting generic cloud-based tools, such as DropBox\footnote{http://dropbox.com} and GoogleDrive\footnote{https://drive.google.com}, or specialized data repositories like DataVerse \cite{dataverse} and Figshare \cite{figshare} to share the results of their research with other scientists. 

Both the guidelines set by journals and the adoption of tools for exchanging investigation materials constitute a first step towards enabling sharing of research results. Yet, they are by no means sufficient. Critical experimental context information regarding the hypothesis investigated and its relation to related work, how research results were concluded from the input sources and how the participants were involved in the different stages of the experiment remain rarely represented. As a result, a growing number of life scientists are now calling for models and tools that can be used to organize, package and share research investigations in a principled manner to leverage an effective communication through the collaboration of research investigation findings\footnote{http://www.force11.org/beyondthepdf2}.

In order to address the above shortcoming, the concept of Research Object (RO) was proposed as an abstraction for sharing research investigation results\cite{bechhofer2010research}. 
In \cite{DBLP:journals/fgcs/BechhoferBRMABCCDDGMONSG13} Bechhofer et al. further provide the vision of Research Objects and their potential role in facilitating sharing, reuse and enabling the reproducibility of scientific investigation results. In this article, we present a realization of this vision using standard Semantic Web technologies. Our main contributions are:
\begin{itemize}
\item
\emph{A core Research Object ontology} which provides a lightweight container structure to encapsulate scientific context information and metadata about research results.
\item
\emph{Extension modules} which show how to adapt the core ontology to  capture domain specific experimental context, including the evolution of the experiment and its relationship to other experiments, the experiment design and the experiment execution settings and traces.
\item
\emph{A family of tools} which allow exploring, sharing and browsing Research Objects and their metadata.
\end{itemize}

The design of Research Objects has been largely driven by the needs for enhancing the knowledge communication between information providers and consumers, in the context of scholarly communication or scientific collaborations. These needs gave rise to a set of principles for Research Objects. To support these principles we find a strong need to ground the representation of Research Objects upon semantic web technologies, so that we can uniquely identify Research Objects, and enable the dissemination and sharing of them using standard protocols and the Web platform. These principles also provide guidance for our implementation of the ontologies, defining the terms and various modules required. 

The Research Object ontologies fill in an important void in enabling a new scholarly publication mode by exposing the actual research asset in structured format and providing semantic explicit descriptions about them and their relationships, for easier reuse and validation. The design of the Research Object ontologies has been driven by the principles of being lightweight, domain-neutral, and extensible.


The rest of the article is organized as follows. In Section 2 we present an updated list of Research Object design principles, based on the practical implementation and the experience we have had since the introduction of its first manifesto\cite{DBLP:journals/fgcs/BechhoferBRMABCCDDGMONSG13}. We then analyze related ontologies and approaches for metadata and experiment representation, highlighting those being reused in the Research Object ontologies (in Section 3). In Section 4, we present a real world scenario from Genome Wide Association Studies, which we use as a running example for illustrating how the RO ontologies can be used to systematically enable the packaging, dissemination and reuse of investigation results. In Section 5, we present the Research Object ontologies, which includes an overview, the Research Object core ontology and its extensions. We also report on the tools that we developed to support the Research Object life cycle in Section 6. Finally, we conclude the article in Section 7.

%% file: ro-khalid.tex
\section{Research Object Principles} \label{section:ro}

Research Objects are treated as first class citizen structures that aggregate  resources in a principled manner. These resources are used and produced during research investigations with the purpose of facilitating the interpretation of scientific findings, enabling their reuse and inspection for reproducibility purposes \cite{belhajjame12:citizens}. Research Objects can accommodate any kind of resources, as long as they are used to provide contextual information of the experiment being described. That said, the design of the Research Object ontologies was driven and influenced by principles based on the systematic analysis of requirements expressed by scientists from the Life Sciences and Astronomy fields, in combination with the ability to cite any of the resources in the Research Object context. Each of these principles is further described below.

\subsection{Preserving Data and Methods}
While the topic of data preservation and sharing received as much attention in academia as in industry, the sharing of code and computational methods has often been neglected (due to private licenses, patents, not being required to publish software, etc.). However, software code or computational methods contain critical information for understanding the exact computational procedure that was used to generate the research findings. Without the software, reproducing computational experiments would be much more expensive to achieve. We argue that a detailed description of the methods and the experimental materials used to process and generate the research datasets \change{must be accessible, just as the research data itself.}


\subsection{Overcoming Obfuscation through Annotation}
Providing scientists with the materials used in the investigation (e.g., datasets and methods) is not enough for making the investigation accessible, interpretable, re-usable and reproducible. In addition, such materials need to be accompanied with annotations (descriptions) specifying contextual information about the hypothesis of the investigation, the roles played by existing datasets, the different steps of the methods (experiments), and the links between experimental results and the conclusions derived by the scientists.

\subsection{Treating the Research Object as a Container}

\change{In order to accommodate data, methods, and the various essential annotations, we need a flexible container structure that allows encapsulating all the different types of information objects. This requirement strongly drives the design of Research Objects, keeping the core structure as open as possible. Furthermore, the Research Object focuses on describing the relationships between the resources aggregated within them, rather than describing the resources themselves\cite{DBLP:journals/fgcs/BechhoferBRMABCCDDGMONSG13}.}

\subsection{Citations and Credit in Research Objects}
It is widely recognized that credit and attribution encourage sharing. Traditionally, this has been, to a large extent, confined to \change{scholarly citations and related citation indexes}. To promote the sharing of Research Objects, scientists should be able to cite and credit other kinds of material on top of which their own research is built on. Therefore, Research Objects should be citable and referable, and built-in mechanisms must be provided to support attribution, citation and credit. Referenceability of Research Objects must happen at both a coarse-grained level, e.g., to cite a Research Object as a whole, and at fine-grained level, e.g., to cite a step in an experiment or a data item within a dataset.

\subsection{Treating Research Objects as Software}
\change{A Research Object can be created or used and released to third parties to communicate findings from ongoing research. A scientific investigation often goes through different hypotheses and designs, in an iterative and dynamic manner. Research Objects that capture ongoing scientific research findings must be managed as a living growing object. A Research Object constitutes a scholarly publication which captures everything that is needed to understand and reproduce the reported findings. 
Therefore, the life cycle of a Research Object is not dissimilar to that of a piece of software, which goes through the cycle of development, testing, release/production, and support. Similarly, the life cycles of Research Objects must also be carefully managed, to ensure that} the evolution of scientific thinking is captured (at least partly).

%% file: related_work.tex
\section{Related Work}
\label{sec:relatedwork}

As stated, Research Objects aim to provide a structured container for describing the context, contents and outcomes of a scientific publication. In this section we describe other existing approaches for describing scholarly publications at different granularity and specificity, highlighting similarities and differences with our work. 




\subsection{Nanopublications and Micropublication}
A nanopublication assertion \cite{groth2010anatomy} encapsulates key findings of a scholarly article as triples and provides attribution to the authors. Micropublications can be thought of as an enrichment of the nanopublication proposal \cite{micropublication}. Using nanopublications, a scientific claim is expressed as an annotated triple, which is primarily extracted from databases. Micropublications augment nanopublications to cover claims made using natural language by researchers in their scholarly articles.

Nanopublications and micropublications provide fine-grained units for representing scientific claims and annotating them with provenance information that associates claims with evidence and other claims. In contrast, Research Objects provide a coarse-grained unit of publication and sharing of scientific knowledge. They are meant to encapsulate elements that are relevant for an entire investigation, as opposed to individual claims. Therefore, Research Objects complement nanopublications and micropublications by providing structured containers for publishing and sharing scientific investigations. Research Objects do not only comprise scientific claims, but allow relating the different parts of an experiment to each other.

\subsection{ReproZip}
ReproZip \cite{reprozip} is a tool that allows authors to systematically capture the provenance of their experiment runs by tracking operating systems calls. This information can then be used to repeat and reproduce the experiment. The scope of Research Objects is wider than the one targeted by tools such as ReproZip, in the sense that Research Objects aim to provide a bigger picture about the investigation carried out, the hypothesis investigated, and the elements that are necessary for understanding both the experiment specifications and their runs.

\subsection{Scientific Publication Packages}
The previous approaches provide the means for capturing specific aspects of a research investigation, e.g., claims and their association with evidence in the case of nanopublications and micropublications and provenance of experiment runs in the case of reproZip. Research Objects, on the other hand, provide a framework for specifying the elements that are necessary for understanding and preserving research investigations as a whole. In this respect, Research Objects are similar to an earlier work by Hunter on Scientific Publication Packages (SPPs) \cite{SPPs}. SPPs involve the encapsulation of raw data, derived products, algorithms, software and textual publications within a container. A Research Object goes beyond SPP on two fronts. First, in Research Objects annotations are first class citizens that are used to ensure that the elements of a Research Object can be interpreted, reused and re-purposed more easily by third-parties. Second, the Research Object ontologies are built on standards and  well established ontologies, and are accompanied with production tools in order to encourage their adoption within the community. 

\subsection{Investigation Study Assay (ISA)}
In the Life Sciences domain there are a number active community standards for describing experimental protocols and context, like the Ontology for Biomedical Investigations (OBI)~\cite{brinkman2010modeling}, and the Investigation/Study/Assay (ISA) format~\cite{rocca2010isa}. Both community standards were motivated by the need from microarray gene expression studies to provide a structured description of gene expression experimentations in order to facilitate knowledge exchange and data integration. Unlike these efforts, the Research Object ontologies are aimed to be domain neutral and their flexible structure can accommodate any domain-specific descriptions about the aggregated resources in a Research Object. This is demonstrated by the workflow-specific extension of the core Research Object ontology (presented later in this paper). Aligning the Research Object ontologies with the ISA model is work in progress.

%% file: example-scenario.tex
\section{Running Scenario: Genome Wide Association Studies}\label{section:example}

\begin{figure}[th!]
	\centerline{\psfig{figure=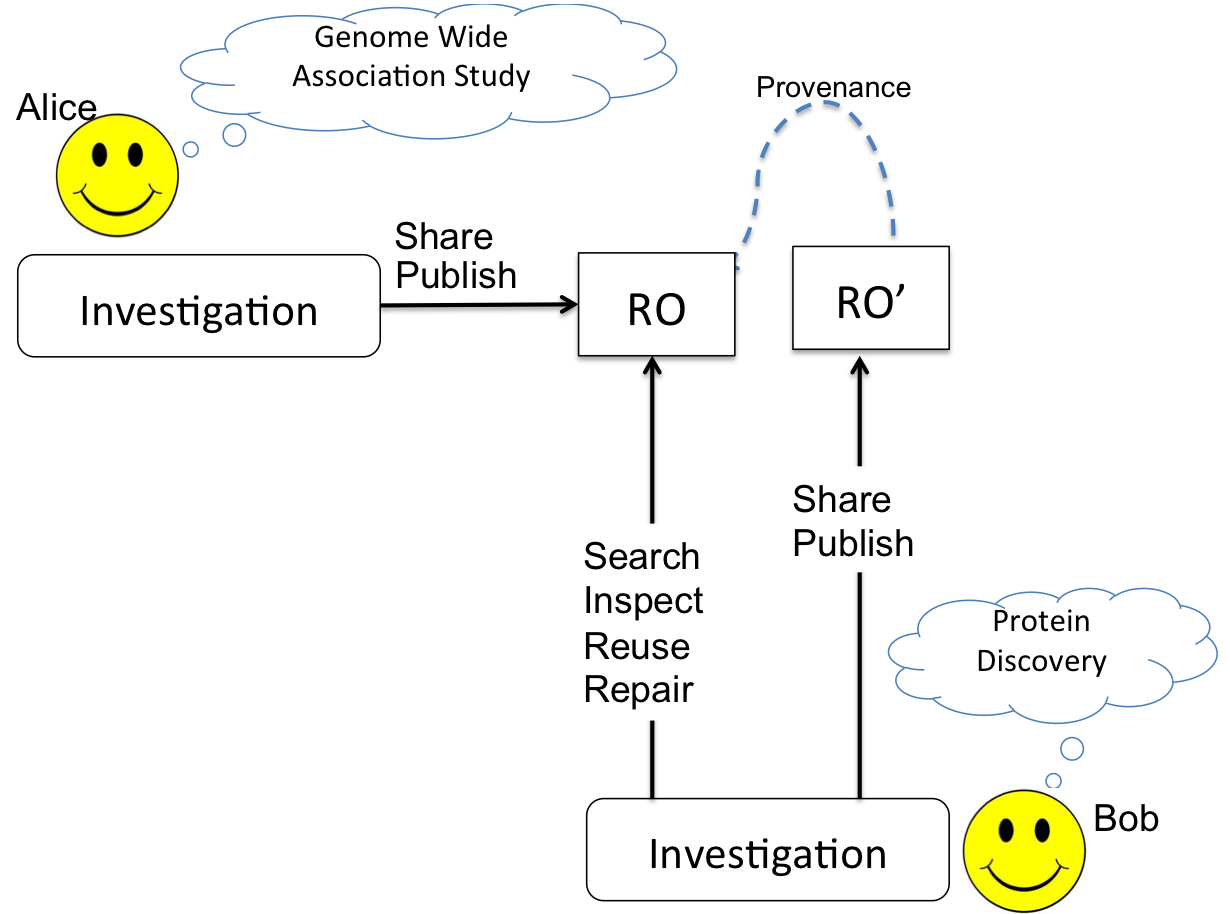,width=90mm} }
	\caption{An example scenario for Research Objects.}
	\label{fig:scenario}
	\end{figure}

In order to illustrate the requirements reported on section \ref{section:ro} and our proposed solutions, we describe in this section the example depicted in Figure~\ref{fig:scenario}. The example illustrates a scenario in which the use of Research Objects is expected to leverage the difficulty of knowledge communication in a collaborative computational research investigation.

In the scenario, Maria is a computational geneticist interested in genome wide association studies. The studies diagnose the possible causes of genetic diseases by identifying mutation genes in the human genome. Given the wealth of genomic knowledge available in public databases, Maria decides to use in silico analysis for helping her to quickly identify those special genes contributing to the development of the rare genetic disease in her study. In order to achieve this goal, Maria adopts scientific workflows as her computational instrument. Scientific workflows provide her with web-based services to build her investigation quickly without writing too much code. 

\subsection{Packing, Publishing and Sharing}
The outcome of Maria's analysis shows a set of genes that have not been identified in any existing public sources as being associated with her studied disease. She wants to share these genes with her collaborators in another laboratory, so that they can verify her results and carry out further web-lab or clinical studies. To help them understand her findings, she includes additional information like provenance information detailing the services used, intermediate results and final results generated by the workflows, background information and interpretation of the results. Maria collects together thsse data as a Research Object and makes it available on the Web.

Once Maria is ready to publish her research findings in an article, she revises her Research Object. Maria is able to include more updated information corresponding to her published results and links to the URI of this Research Object. This way, anyone reading her articles can also benefit from the access to a complete package of additional information that helps understanding and verifying Maria's results.

\subsection{Reusing, Reproducing and Re-publishing}
Bob is another computational geneticist interested in discovering proteins related to mutation genes that contributed to the same genetic disease being studied by Maria. After reading Maria's publication, he believes that her set of genes would be very helpful for him to expand proteomics understanding about this genetic disease. Since Maria has shared the resources that are required for him to reproduce or re-execute the workflow, including the link to the platform and the original experiment settings, Bob is able to quickly verify Maria's experiment process and start from her findings. He expands Maria's original workflow to include queries to public proteomics data sources and services. When he is ready to publish his workflow and findings, instead of simply citing Maria's original publication, he also points his new Research Object to Maria's Research Object, providing additional attribution and tracking the provenance of these new findings in a structured way.

In the following sections, we show how the Research Object ontologies and Semantic Web technologies can be used to represent the Research Object that Maria created and shared and track the relationship between Bob's Research Object and Maria's Research Object.

%% file: ro_ontologies_overview.tex
\section{The Research Object Ontologies}
\label{sec:ontologies_overview}
The Research Object Ontologies were developed after a systematic analysis of the requirements introduced in Section \ref{section:ro}. As a result, the ontologies are divided in a core ontology (which provides the basic means to create Research Objects) and different extension modules which further extend and describe the core. In this section we first provide an overview of the ontologies in subsection \ref{sub:overview}, we describe the vocabularies and standards on top of which the Research Object ontologies are built in subsection \ref{sec:underlyingvocabularies} and we finally introduce the core and its extensions in subsections \ref{sub:core} and \ref{sec:extension_ontologies}.

\subsection{Overview}\label{sub:overview}

\begin{figure*}[ht]
  \centering
  \includegraphics[width=0.9\textwidth]{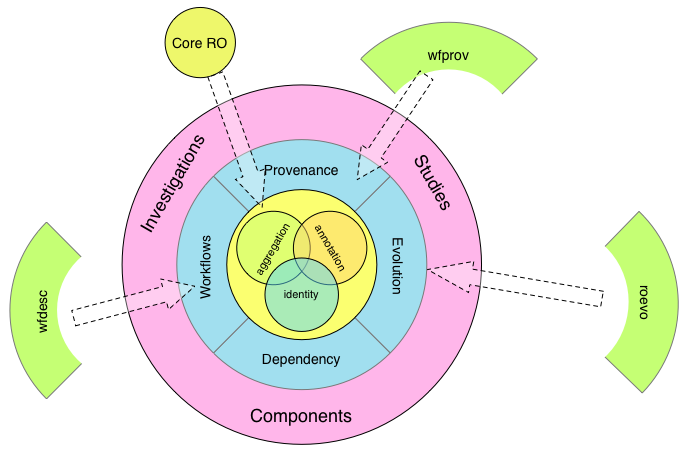}
  \caption{Overview of the Research Object Model with the different ontologies that encode it: Core RO for describing the basic Research Object structure, roevo for tracking Research Object evolution and wfprov and wfdesc for describing workflows and capturing the provenance traces of their execution.}
  \label{fig:wm_abstract}
\end{figure*}

Figure \ref{fig:wm_abstract} shows an overview of the Research Object model, distinguishing between core and extension modules. The design of the core is driven by three main requirements, namely a mechanism for uniquely identifying Research Objects, a means for aggregating resources within a Research Object, and the ability to annotate the Research Object, its constituent resources and their relationships. The extended modules address the need for defining dependencies between Research Objects, the Research Object evolution over time and the specification of scientific workflows (experiments) and provenance traces of their executions, which are very common in any computational experiment.

As illustrated in Figure \ref{fig:wm_abstract}, we have encoded the Research Object model using a family of ontologies. The Research Object ontologies have been developed in an incremental manner, taking feedback from various users and collaborators at every iteration. This incremental development approach ensured that the ontologies developed cater for the concrete requirements specified by the users of the ontologies. Requirements have been gathered and documented online\footnote{\url{http://www.wf4ever-project.org/wiki/display/docs/Requirements}}. Every release of the ontologies is tracked in the project Github repository\footnote{\url{https://github.com/wf4ever/ro}}. 

We have striven to reuse terms from existing popular vocabularies for interoperability and extended them whenever necessary to cater for our requirements. Whenever the extension was not necessary, we directly reused the existing vocabularies. For example, no new terms have been introduced for describing dependencies, investigations, components and studies, since there exist ontologies that implement these aspects, e.g., the Dublin Core vocabulary\footnote{\url{http://dublincore.org}} provides the means for specifying dependencies, and the ISA model\footnote{\url{http://www.isa-tools.org}} provides concepts for specifying investigations and studies.




We have favored minimality over completeness, making sure that the Research Object ontologies contain extension points for users to customize them. This way third parties are able to define specific types of Research Objects or to describe more detailed information that is specific to their tools or application scenarios. In this respect, we show  in section \ref{sub:wf-ro} how a specific kind of Research Objects that describe research findings from scientific workflows can be represented. Similarly, users of the Research Object ontologies might want to be able to define research findings generated by other types of computational instruments, like simulation libraries, workflow optimization tools, etc. 

\subsection{Research Object Underlining Vocabularies}
\label{sec:underlyingvocabularies}
The Research Object ontologies have been developed on top of well used vocabularies and standards. In particular, the Research Object Core Ontology reuses a well stablished vocabulary for describing aggregations (OAI-ORE) and the Annotation Ontology (AO) for specifying annotations. On the other hand, the extension modules reuse the W3C Provenance standard model (PROV) for keeping track of the results of an experiment and the evolution of Research Objects. These reused vocabularies vocabularies are further described below. 

\paragraph*{Open Archive Initiative - Object Exchange and Reuse (OAI-ORE)}
The Object Exchange and Reuse (ORE) model\footnote{\url{http://www.openarchives.org/ore/1.0/toc.html}} is a community standard developed by the Open Archive Initiative (OAI) to facilitate interoperable descriptions and exchange of aggregations of web resources. ORE defines a lightweight aggregation structure and provides a basis for the Research Object ontologies. However, it does not provide terms for the annotation or description of Research Objects or its aggregated resources. To cater for this requirement, we borrow strength from another community vocabulary, the Annotation Ontology.

\paragraph*{Annotation Ontology and Open Annotation Model}
Annotations are used to describe the Research Object, its constituent resources and the relationships between resources. In order to keep the Research Object ontology as domain neutral as possible, we adopted an annotation framework that can accommodate annotations expressed by any specific vocabularies, like Dublin Core\footnote{\url{http://dublincore.org}}, Friend of a Friend (FOAF)\footnote{\url{http://www.foaf-project.org}}, etc. We had two well-developed annotation vocabularies to consider, the Annotation Ontology (AO)\footnote{\url{http://code.google.com/p/annotation-ontology}} and the Open Annotation Model\footnote{\url{http://www.openannotation.org/spec/beta}}. Both ontologies share similar structure, but the AO has explicit support for expressing the provenance and attribution of the annotations by using the Provenance, Authoring and Versioning ontology \cite{pav}. This is particularly important considering that a Research Object could be the result of a collaboration between investigators, containing information contributed by diverse scientists.


\paragraph*{PROV-O}
Research Objects should provide provenance information about themselves and the results of the experiments aggregated in them. Thus, we have designed two ontologies for this purpose. The first one is an ontology developed for tracking Research Object evolution (roevo) and the second one is a provenance ontology for recording scientific workflow executions and their results step by step (wfprov). Both ontologies are built extending the W3C standard Provenance Ontology (PROV-O) \cite{w3c-prov-o}, a W3C recommendation for facilitating the interoperability and exchange of provenance information between systems and applications on the Web. In PROV-O, resources (prov:Entities) are used or generated by activities (prov:Activities) which may had a responsible agent (prov:Agent). By keeping track of the chain of activities that use and generate entities, we are able to record the evolution of Research Objects and how the results of the experiments are derived from the inputs step by step.





\begin{figure*}[ht!]
	\centering
\includegraphics[width=0.7\textwidth]{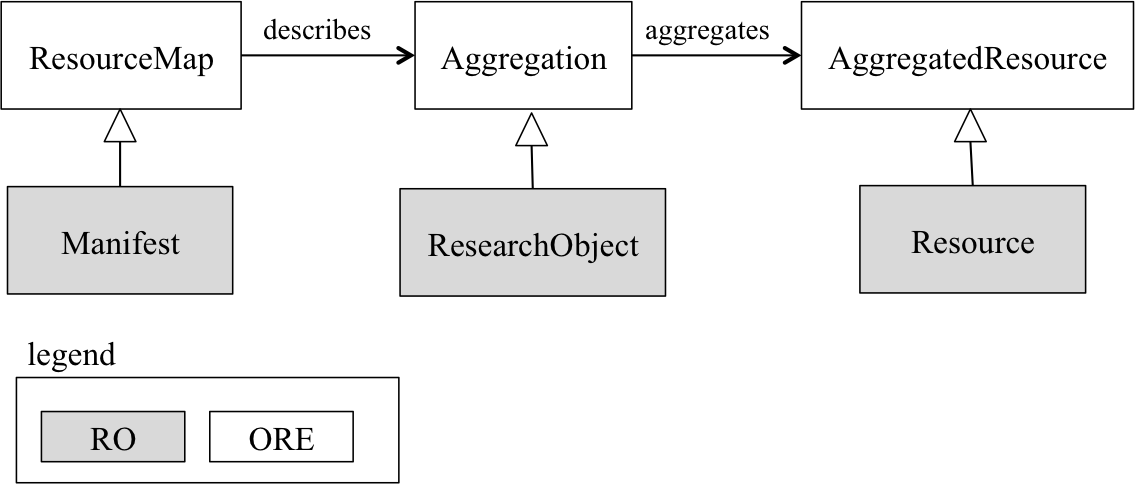} 
	\caption{Research Object as an ORE aggregation.}
	\label{fig:ro_ontology}
  \end{figure*} 

\subsection{The Research Object Core Ontology}\label{sub:core}
We present in this section the Research Object Core ontology, discussing its design decissions and how it extends the ORE vocabulary for aggregating resources (in subsection \ref{sub:aggregations}) and the AO vocabulary for enabling annotations (in subsection \ref{sub:annot}).

\subsubsection{Describing Aggregation Structures}\label{sub:aggregations}
To cater for the specification of aggregation structures, we built the Research Object Core Ontology\footnote{\url{ro:http://purl.org/net/wf4ever/ro#}} upon the popular ORE vocabulary\footnote{ \url{ore:http://www.openarchives.org/ore/terms/}}. ORE defines a standard for the description and exchange of aggregations of Web resources. Research Objects are defined in terms of three main ORE concepts:
\begin{itemize}
\sloppy
\item \texttt{ore:Aggregation}, which groups together a set of resources so that it can be treated as a single resource. 
\sloppy
\item \texttt{ore:AggregatedResource}, which refers to a resource aggregated in an \texttt{ore:Aggregation}. An \texttt{ore:AggregatedResource} can be aggregated by one or more \texttt{ore:Aggregation} and it does not have to be physically included in an \texttt{ore:Aggregation}. An \texttt{ore:Aggregation} can aggregate other \texttt{ore:Aggregations}. 
\sloppy
\item \texttt{ore:ResourceMap}, which is a resource that provides descriptions of an \texttt{ore:Aggregation}.
\end{itemize}

Using ORE, a Research Object can be defined as \change{an aggregation} that contains other resources, \change{such as} workflows specifying experiments, provenance logs, other \change{information} objects and annotations. \change{Therefore, the ORE vocabulary provides the perfect foundation for us to define the Research Object container structure.} Figure \ref{fig:ro_ontology} illustrates the main terms that constitute the Research Object Core Ontology, which we describe below.

\begin{itemize}
\item
\texttt{ro:ResearchObject}, represents an aggregation of resources. It is a sub-class of \texttt{ore:Aggregation} and acts as an entry point to the Research Object.
\item
\texttt{ro:Resource}, represents a resource that can be aggregated within a Research Object and is a sub-class of \texttt{ore:AggregatedResource}. A \texttt{ro:Resource} can be a \texttt{ro:Dataset}, \texttt{ro:Paper}, \texttt{ro:Software} or \texttt{ro:Annotation}. Typically, a \texttt{ro:ResearchObject} aggregates multiple \texttt{ro:Resources}, specified using the property \texttt{ore:aggregates}.
\item
\texttt{ro:Manifest}, a sub-class of \texttt{ore:ResourceMap}, represents a resource that is used to describe a \texttt{ro:ResearchObject}. It plays a similar role to the manifest in a JAR or a ZIP file, and is primarily used to list the resources that are aggregated within the Research Object.
\end{itemize}

\subsubsection{Enabling Annotations}\label{sub:annot}
To properly record annotations of Research Objects, we selected the Annotation Ontology (AO) release 2.0b2~\cite{COG11}. The Research Object Core ontology  reuses three main Annotation Ontology concepts for defining annotations:  \texttt{ao:Annotation}\footnote{\url{ao:http://purl.org/ao/}}, used for representing the annotation itself; \texttt{ao:Target}, used for specifying the \texttt{ro:Resource}(s) or \texttt{ro:ResearchObject}(s) subject to annotation; and \texttt{ao:Body}, which comprises a description of the target.

Research Objects use annotations as a means for decorating a resource (or a set of resources) with metadata information. The body is specified in the form of a set of RDF statements, which can be used to annotate  the date of creation of the target, its relationship with other resources or Research Objects, etc. Also, annotations can be provided for human consumption (e.g. a description of a hypothesis that is tested by a workflow-based experiment), or for machine consumption (e.g. a structured description of the provenance of results generated by a workflow run). Both kinds of annotations are accommodated using Annotation Ontology structures.

As an example, Listing \ref{lst:examplero} shows how a Research Object with title "GWAS to Kegg" can be specified using an aggregation and annotations. The listing describes the basic elements of the Research Object that Maria shares with her collaborator. In this case, the Research Object includes the inputs (\texttt{<data2.csv>}), workflows (\texttt{<workflow34.xml>}), hypothesis (\texttt{<hypothesis.txt>}) and provenance record of her experiment (in a \texttt{<provenance.rdf>} file), as well as metadata identifying the creator and the date of creation. Additional attribution information (contributors, publishers, license information, etc.) could also be added if needed to properly reference the Research Object or any of its parts.

\begin{lstlisting}[
  language=RDF_N3,
  float=b!,
  label= lst:examplero,
  caption={An example of the core elements of a Research Object}
]
@base <http://example.com/ro/389/> .
@prefix dct: <http://purl.org/dc/terms/> .
@prefix ore: <http://www.openarchives.org/ore/terms/> .
@prefix ao: <http://purl.org/ao/> .
@prefix ro: <http://purl.org/wf4ever/ro#> .
@prefix roterms: <http://purl.org/wf4ever/roterms#> .
@prefix xsd: <http://www.w3.org/2001/XMLSchema#> .

<> a ro:ResearchObject ;
 dct:title      "GWAS to kegg" ;
 dct:creator    </foaf/maria> ;
 ore:aggregates <workflow34.xml>, <data2.csv>, <provenance.rdf>, <hypothesis.txt>, <#annotation2>, <#annotation3> .

<workflow34.xml> a ro:Resource .
<data2.csv> a ro:Resource .
<provenance.rdf> a ro:Resource .
<hypothesis.txt> a roterms:Hypothesis .

<#annotation2> a ro:SemanticAnnotation ;
 ro:annotatesAggregatedResource <workflow34.xml> ;
 ao:body <workflow34.wfdesc.ttl> ;
 dct:created "2013-02-12T19:39:29.379Z"^^xsd:dateTime .

<#annotation3> a ro:SemanticAnnotation ;
 ro:annotatesAggregatedResource <provenance.rdf> ;
 ao:body <run-481.wfprov.ttl> ;
 dct:created "2013-02-15T19:41:12.792Z"^^xsd:dateTime .
 
\end{lstlisting}

%% file: ontologies.tex
\subsection{Research Object Extension Ontologies}
\label{sec:extension_ontologies}

The core Research Object ontology presented in the previous section is a general purpose ontology. The core Research Object Ontology can be extended in many ways depending on our domain specific requirements. In this section we present two  extensions to the core Research Object ontology. The first one (described in the subsection \ref{subsec:roevo} shows how to specify the evolution of Research Objects over time, while the second one focus on describing the methods of experiments and the traces of their executions in section \ref{sub:wf-ro}. Just as happened with the core Research Object ontology, extension ontologies are built on existing vocabularies and standards \change{as much as possible}. In particular, we extend W3C PROV-O \cite{w3c-prov-o} and use Dublin Core terms\footnote{\url{http://dublincore.org/2010/10/11/dcterms}} and Friend of A Friend (FOAF)\footnote{\url{http://www.foaf-project.org}} vocabularies for capturing additional metadata.

\subsubsection{Tracking Research Object evolution using the \textit{roevo} Ontology}
\label{subsec:roevo}
The \textit{roevo} ontology is an extension to the core Research Object ontology for describing the life cycle of Research Objects.
In order to achieve this goal, we need to track and describe the changes made to a Research Object at different levels of granularity: the changes made to a  Research Object as a whole (its creation and current status) and the changes made to the individual resources aggregated within the Research Object (additions, modifications and removals). We aim to provide sufficient details so as to be able to roll back to a particular version and to perform quality control over the changes. Therefore, we need to describe when a change took place, who performed the change, and record the dependency relationships between the different changes. 
A change is closely related to the provenance of a particular version of a Research Object or a resource. The PROV-O ontology provides all the foundational information elements for us to build the evolution ontology, so it is used as a baseline and extended appropriately.

\begin{figure*}[ht]
  \centering
  \includegraphics[width=0.8\textwidth]{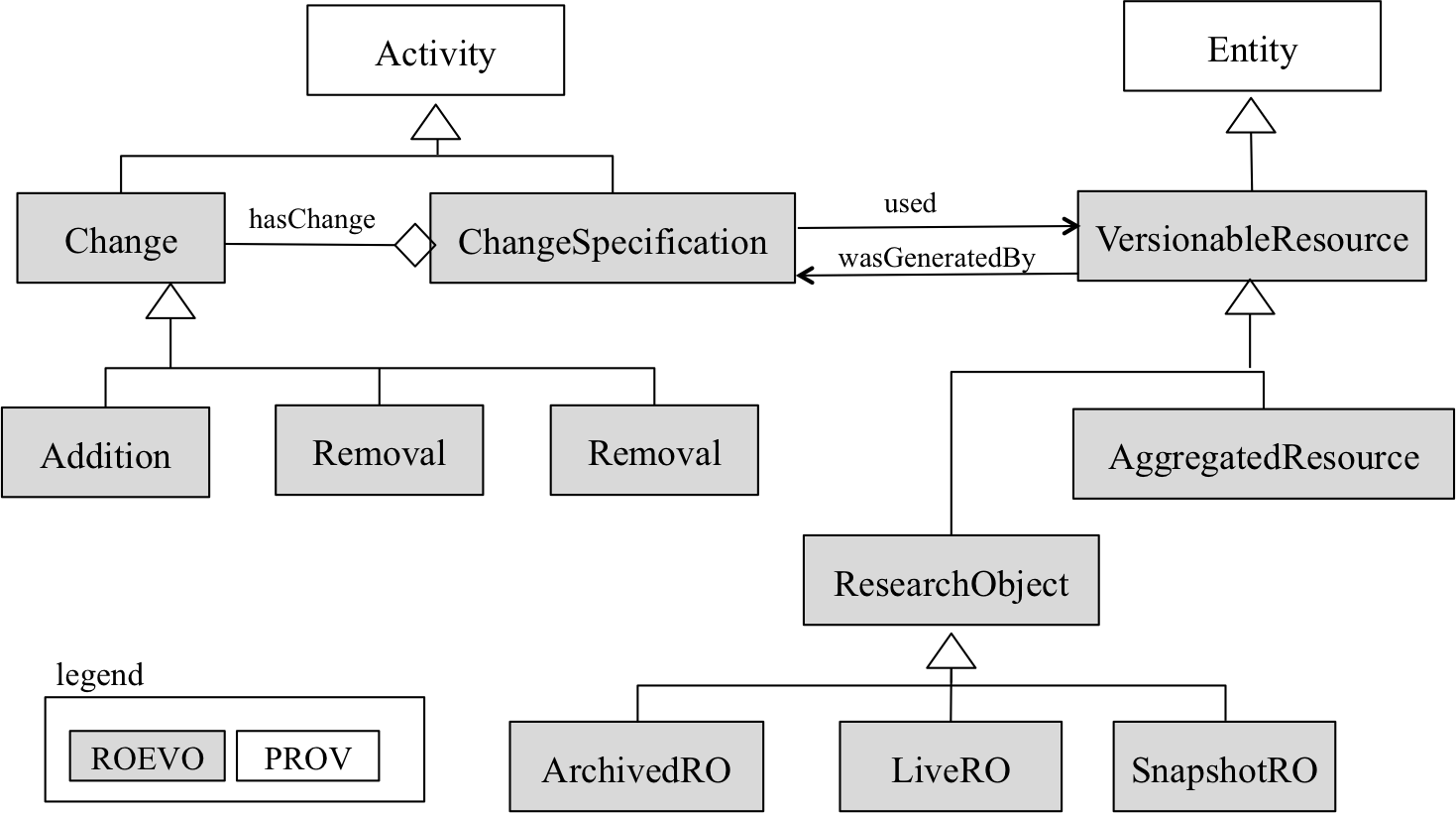}
  \caption{The \textit{roevo} ontology extending PROV-O terms.}
  \label{fig:ro_evo}
\end{figure*}

Figure~\ref{fig:ro_evo} illustrates the core concepts of the \textit{roevo} ontology and how it extends the PROV-O:
\begin{itemize}
\item Three sub-classes of \texttt{ro:ResearchObject} have been created to capture different states of a Research Object during its life time. A \texttt{roevo:LiveRO} is a Research Object designed to capture research findings during a live investigation. A Live Research Object can be changed, archived or snapshotted. A \texttt{roevo:ArchivedRO} can be seen as a production Research Object to be preserved and archived, such as the one describing findings published in an article, and it can no longer be changed. Finally, a \texttt{roevo:SnapshotRO} represents a live Research Object at a particular time (a particular version of a live investigation).
\item Both a snapshot of a live Research Object and an archived Research Object can be regarded as a versioned Research Object, i.e. a \texttt{roevo:VersionableResource}. Since we want to track the provenance of any \texttt{roevo:Versionable- Resource}, we consider this class a subtype of \texttt{prov:Entity}. We then reuse PROV-O properties (\texttt{prov:used} and  \texttt{prov:wasGeneratedBy}) to describe the provenance of all the changes made to this entity, pointing to the activity that generated the changes, to the source Research Object from which the current version was derived, and all the agents involved in the changes.
\item A change is a type of \texttt{prov:Activity}, which means that it has a start time, an end time, an input entity and a resulting entity. Also a change leading to a new Research Object can constitute a series of changes. Therefore, we have a composite \texttt{roevo:ChangeSpecification} activity, which has a number of unit \texttt{roevo:Change}s. A unit change can be adding, removing or modifying a resource or a Research Object. But these different changes share the same pattern of taking an input entity and producing an output entity, which can all be covered by properties from PROV-O.
\end{itemize}

As an example, Listing \ref{lst:roevo_example} specifies that {\tt ro-2} is a Research Object that underwent a change specification that consisted in the addition of two resources: an annotation and  a conclusion. It also specifies that {\tt ro-2} is a revision of a snapshot of another Research Object {\tt ro-1} and that  {\tt ro-2}  was archived.

\begin{lstlisting}[language=RDF_N3,label=lst:roevo_example,
  caption={Using the \textit{roevo} ontology to describe the snapshots of a Research Object}]
@base <http://example.com/ro/> .
@prefix xsd:     <http://www.w3.org/2001/XMLSchema#> .
@prefix roevo: <http:1//purl.org/wf4ever/roevo#> .
@prefix prov: <http://www.w3.org/ns/prov#> .

<ro-2/change_specification/1>
  a roevo:ChangeSpecification ;
  roevo:hasChange
	<ro-2/change/1> , <ro-2/change/2> .

<ro-2/change/1>
  a roevo:Change , roevo:Addition ;
  roevo:relatedResource
	<ro-2/.ro/annotations/1> .

<ro-2/change/2>
  a roevo:Change , roevo:Addition ;
  roevo:relatedResource
	<ro-2/conclusion.pdf> .

<ro-2/>  
  a roevo:ArchivedRO ;
  roevo:archivedAtTime
	"2013-02-12T19:46:06.677Z"^^xsd:dateTime ;
  roevo:archivedBy
	<http://example.org/public> ;
  roevo:isArchiveOf
	<ro-1/> ;
  roevo:wasChangedBy>
	<ro-2/change_specification/1> ;
  prov:wasRevisionOf
	<ro-1-snapshot/> .
\end{lstlisting}

\begin{figure*}[ht]
  \centering
  \includegraphics[width=0.7\textwidth]{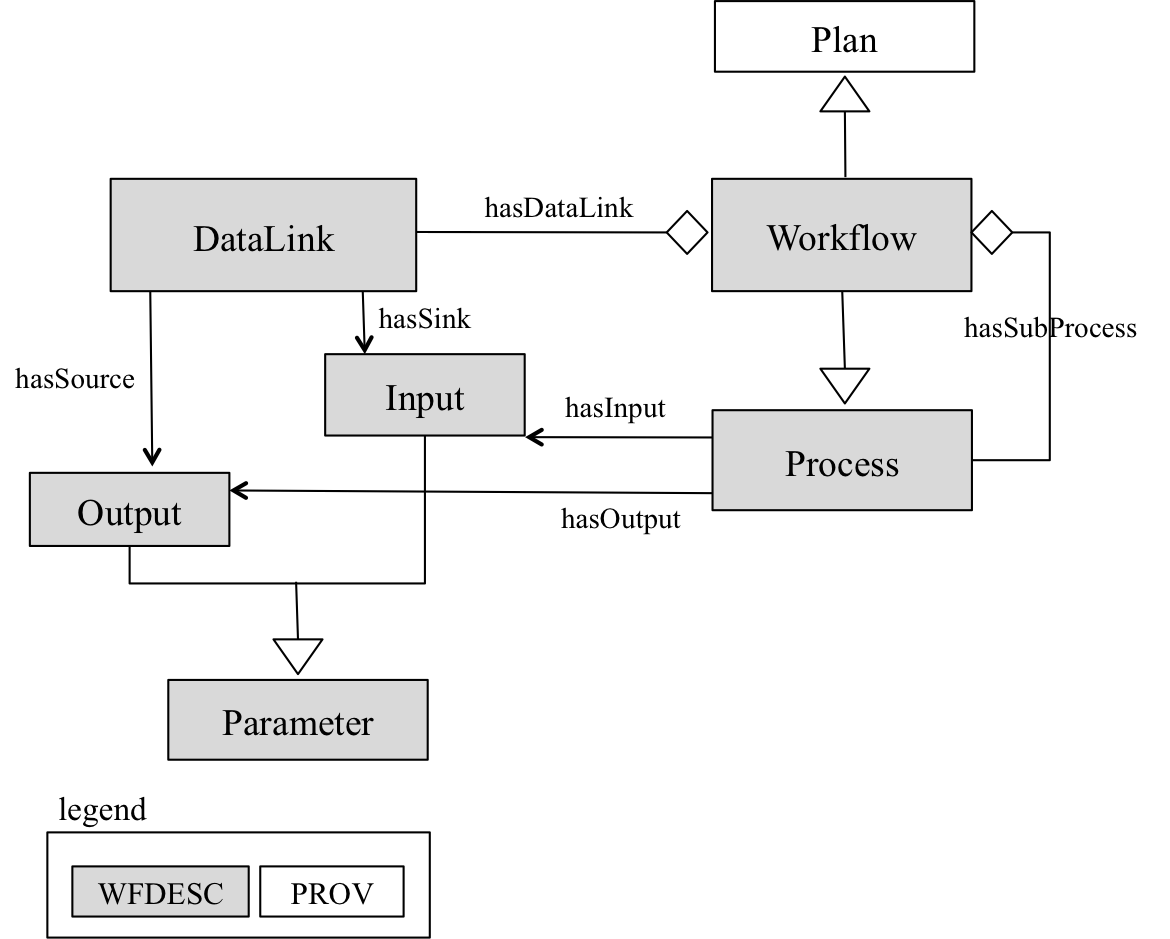}
  \caption{The \textit{wfdesc} ontology and its relation to PROV-O.}
  \label{fig:wfdesc}
\end{figure*}

\subsubsection{Workflow-Centric Research Objects}\label{sub:wf-ro}

Scientific workflows have emerged in the last decade as the technology of choice for modeling experiments and automating their execution \cite{deelman:workflows}. Scientific workflows are often defined as directed acyclic graphs where the nodes correspond to steps in the experiment and the edges specify the flow of data between the steps. Workflows are scholarly valuable resources in the sense that they document the method/experiment followed by the scientists. Moreover, they support unambiguous interpretations of research findings, and  support the verification of their reproducibility.




\change{This section presents two Research Object extension ontologies designed to} support the design of a specific class of Research Objects, which we refer to as {\em Workflow-Centric Research Objects}. Workflow-Centric Research Objects aim to capture the specification of the workflow modeling the experiment as well as all the information about the workflow execution (e.g., its configuration, the datasets  used and generated as a result of the workflow execution, intermediate results, software codes and web services used during the execution, etc.). 
\change{These two aspects are inseparable for describing a Workflow-Centric Research Object: the design of the experiments provides the scientific rationale behind the findings, and the provenance of the execution provides the essential evidence.} 

Both extension ontologies provide only the minimal set of terms that we believe are commonly shared by various existing workflow systems, like Taverna \cite{taverna}, Kepler \cite{kepler}, Vistrails \cite{vistrails} and Wings \cite{wings}. As shown with the Research Object core ontology, the extension ontologies can be further specialized to be accomodated to each workflow system.


\paragraph{Describing Experiment Design using the \textit{wfdesc} Vocabulary\newline}

\change{In order to describe Workflow-Centric Research Objects, the workflow description vocabulary \textit{wfdesc}\footnote{\url{wfdesc: http://purl.org/wf4ever/wfdesc#}} defines several specific concepts and properties that are involved in a workflow specification.} The choice of these concepts was performed by examining the commonalities between major data driven workflow systems, namely Taverna\footnote{\url{http://www.taverna.org.uk}}, Wings\footnote{\url{http://http://wings-workflows.org}} and Galaxy\footnote{\url{http://galaxyproject.org}}, to cite a few.


Figure \ref{fig:wfdesc} illustrates the terms that compose the \textit{wfdesc} ontology. Using such ontology, a workflow is described using the following three main terms:
\begin{itemize}
\item
\texttt{wfdesc:Workflow} refers to a directed acyclic graph in which the nodes are the steps performed in the experiment and the edges represent the data links. It is defined as a subclass of the \texttt{prov:Plan} concept from the PROV-O ontology, which represents a set of actions or steps intended by one or more agents to achieve some goals \cite{w3c-prov-o}. 
\item
\texttt{wfdesc:Process} is used to describe a class of actions that when enacted give rise to process runs. Processes specify the software component (e.g., web service, script) responsible for undertaking those actions.
\item
\texttt{wfdesc:DataLink} is used to encode the data dependencies between the processes that constitute a workflow. Specifically, a data link connects the output of a given process to the input of another process, specifying that the artifacts produced by the former are used as input for the latter.
\end{itemize}

\begin{figure}
\centerline{\psfig{figure=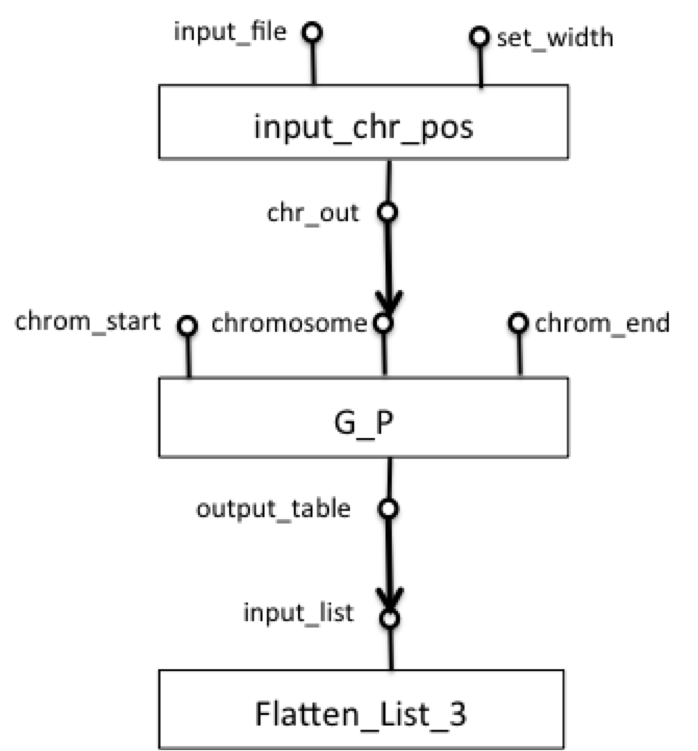,width=60mm} }
\caption{Example workflow.}
\label{fig:example_wf}
\end{figure}

As an example, Listing \ref{lst:wfdesc_example} illustrates how a workflow can be specified using \textit{wfdesc}. The same workflow is depicted in Figure \ref{fig:example_wf}. The workflow in question, labeled {\em mining$\_$the$\_$Kegg$\_$path}, is composed of three processes: {\tt <\#proc/input$\_$chr$\_$pos/>}, {\tt <\#proc/G$\_$P/>} and  {\tt <\#proc/Flatten$\_$List$\_$3/>}. Such processes are connected in sequence using two data links. Notice that the second process  {\tt <\#proc/G$\_$P/>}  has three inputs, two of which {\tt chrom$\_$start} and  {\tt chrom$\_$end} are not connected to any data link. This is because these are configuration parameters that are set before running the workflow. They are set once by the workflow user for multiple workflow runs. 

\begin{lstlisting}[language=RDF_N3,label=lst:wfdesc_example,
  caption={Example of a workflow specified using \textit{wfdesc}}]
@base <http://example.com/ro/389/workflow34.xml> .
@prefix rdfs: <http://www.w3.org/2000/01/rdf-schema#> .
@prefix wfdesc: <http://purl.org/wf4ever/wfdesc#> .

<#> a wfdesc:Workflow ;
  rdfs:label "Mining_the_Kegg_path" ;
  wfdesc:hasWorkflowDefinition <workflow34.xml> ;
  wfdesc:hasSubProcess <#proc/input_chr_pos/>, <#proc/G_P/>, <#proc/Flatten_List_3/> ;
  wfdesc:hasDataLink <#datalink/1>, <#datalink/2> ;
  wfdesc:hasInput <#in/input_file>, <#in/set_width> .

<#proc/G_P/> a wfdesc:Process ;
  rdfs:label "Gene_to_Pathway" ;
  wfdesc:hasInput <#proc/G_P/in/SNP>, <#proc/G_P/in/chromosome>, <#proc/G_P/in/chrom_end>, <#proc/G_P/in/chrom_start> ; 
  wfdesc:hasOutput <#proc/G_P/out/output_table> . 
 
# The output of the proc "input_chr_pos" is the input to the following proc "Gene_to_Pathway" 
<#datalink/1> a wfdesc:DataLink ;
  wfdesc:hasSource <#proc/input_chr_pos/out/chr_out> ;
  wfdesc:hasSink <#proc/G_P/in/chromosome> .

# The output of the proc "Gene_to_Pathway" is the input to the following proc "Flatten_List_3" 
<#datalink/2> a wfdesc:DataLink ;
  wfdesc:hasSource <#proc/G_P/out/output_table> ;
  wfdesc:hasSink <#proc/Flatten_List_3/in/inputlist> .
\end{lstlisting}

\paragraph{Describing Provenance using the \textit{wfprov} Vocabulary\newline}

The \textit{wfprov} ontology is used to describe the provenance traces obtained by executing workflows. As it happened with \textit{roevo}, \textit{wfprov} is also defined as an extension of PROV-O\footnote{reported in the W3C PROV implementation report: \url{http://www.w3.org/TR/prov-implementations/}}


\begin{figure*}[ht]
  \centering
  \includegraphics[width=0.7\textwidth]{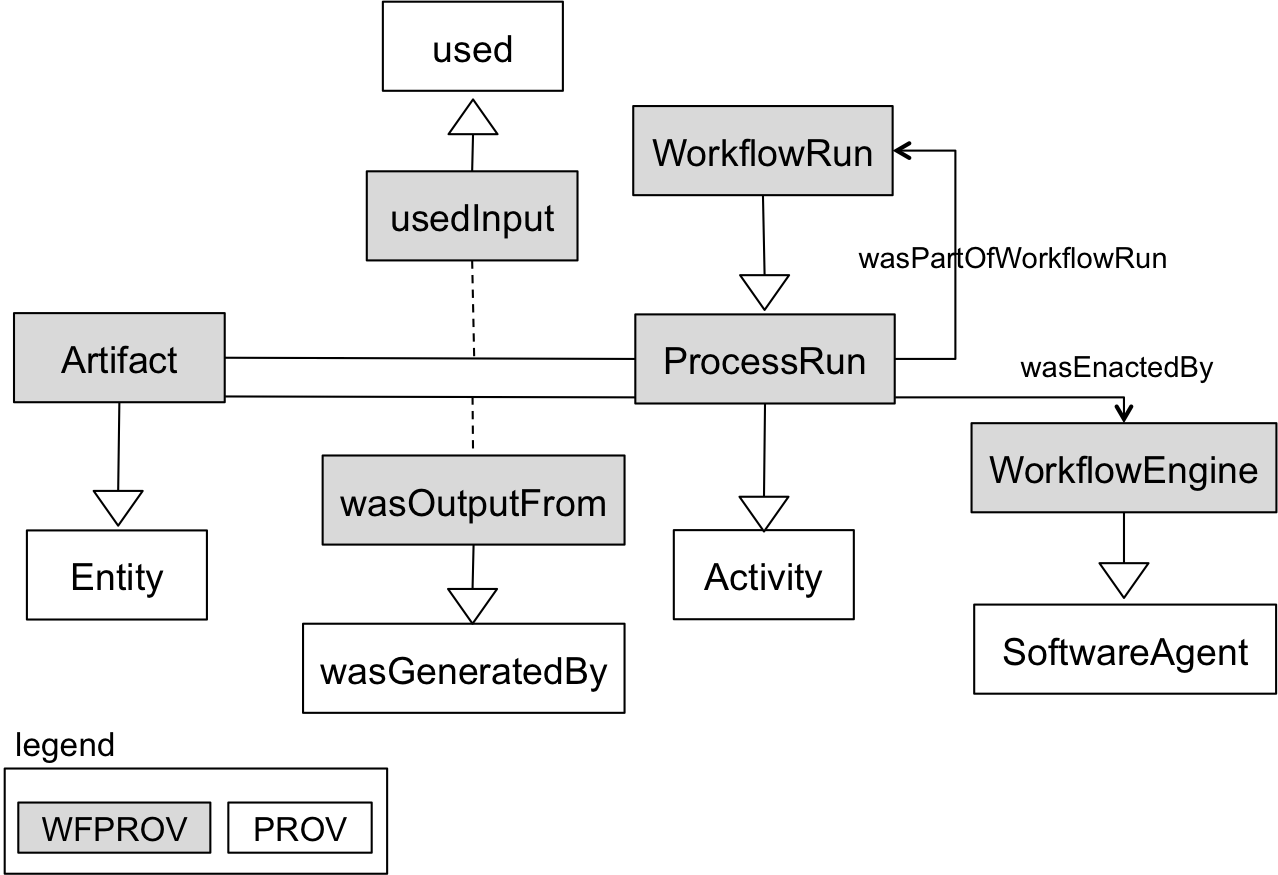}
  \caption{The \textit{wfprov} ontology and its relationship to PROV-O.}
  \label{fig:wfprov}
\end{figure*} 

Figure \ref{fig:wfprov} illustrates the structure of the \textit{wfprov} ontology and its alignments with the PROV-O ontology. A a workflow run~(\texttt{wfprov:WorkflowRun}) represents the enactment of a given workflow. It is composed of a set of process runs~(\texttt{wfprov:Process- Run}), each representing the enactment of a process. A process run may use some artifacts~(\texttt{wfprov:Arti- fact}) as input and generate others as output. A process run is enacted by a workflow engine~(\texttt{wfprov:WorkflowEngine}), which can be seen as a \texttt{prov:SoftwareAgent}.

By chaining the usage and generation of artifacts together, the \textit{wfprov} ontology allows scientists to trace the lineage of workflow results. For example the user can identify the input artifacts that were used to feed the wokflow run (as a whole) to obtain a given output that was generated by the workflow run.

Listing \ref{lst:wfprov_example} specifies a workflow run using \textit{wfprov}. The listing specifies that a workflow run (\texttt{<\#run-481>}) was obtained by enacting the workflow described in Listing \ref{lst:wfdesc_example} (\texttt{<workflow34.xml\#>}) using the Taverna workflow engine. It also specifies the input data (\texttt{<data/input\_file>} and \texttt{<data/set\_width>}) and the output data (\texttt{<data/ G\_P/out/output\_table>}) of the workflow. Moreover, it describes one of the costituent process runs of the workflow run (\texttt{<\#run-481/G\_P>}). The other two process runs were omitted as they are specified in a similar manner.

\begin{lstlisting}[language=RDF_N3,label=lst:wfprov_example,
  caption={Example of a workflow run specified using \textit{wfprov}}]
@base <http://example.com/ro/389/> .
@prefix rdfs: <http://www.w3.org/2000/01/rdf-schema#> .
@prefix foaf: <http://xmlns.com/foaf/0.1/> .
@prefix prov: <http://www.w3.org/ns/prov#> .
@prefix wfprov: <http://purl.org/wf4ever/wfprov#> .

<#run-481> a wfprov:WorkflowRun ;
  rdfs:label "Mining_the_Kegg_path" ;
  wfprov:describedByWorkflow <workflow34.xml#> ;
  prov:startedAtTime "2013-03-15T11:01:54Z"^^xsd:dateTime ;
  prov:endedAtTime "2013-03-15T11:03:12Z"^^xsd:dateTime ;
  wfprov:wasEnactedBy <#taverna> ;
  wfprov:usedInput <data/input_file> ;
  wfprov:usedInput <data/set_width> .

<#taverna> a wfprov:WorkflowEngine ;
  foaf:homepage <http://taverna.org.uk/download/> .

<#run-481/G_P> a wfprov:ProcessRun ;
  rdfs:label "Gene_to_Pathway" ;
  wfprov:wasPartOfWorkflowRun <#run-481> ;
  wfprov:describedByProcess <workflow34.xml#proc/G_P/> ;
  wfprov:usedInput <data/G_P/in/SNP>, <data/G_P/in/chromosome>, <data/G_P/in/chrom_end>, <data/G_P/in/chrom_start> . 
  
<data/G_P/out/output_table> a wfprov:Artifact ;
  prov:generatedAtTime "2012-03-15T11:02:00Z"^^xsd:dateTime ;
  wfprov:wasOutputFrom <#run-481/G_P> .
  
<data/G_P/in/SNP> a wfprov:Artifact ;
  wfprov:describedByParameter <workflow34.xml#proc/G_P/in/SNP> .
\end{lstlisting}


%% file: tools.tex
\section{Research Object Family of Tools}
\label{sec:tools}

\begin{figure*}[ht]
  \centering
  \includegraphics[width=0.85\textwidth]{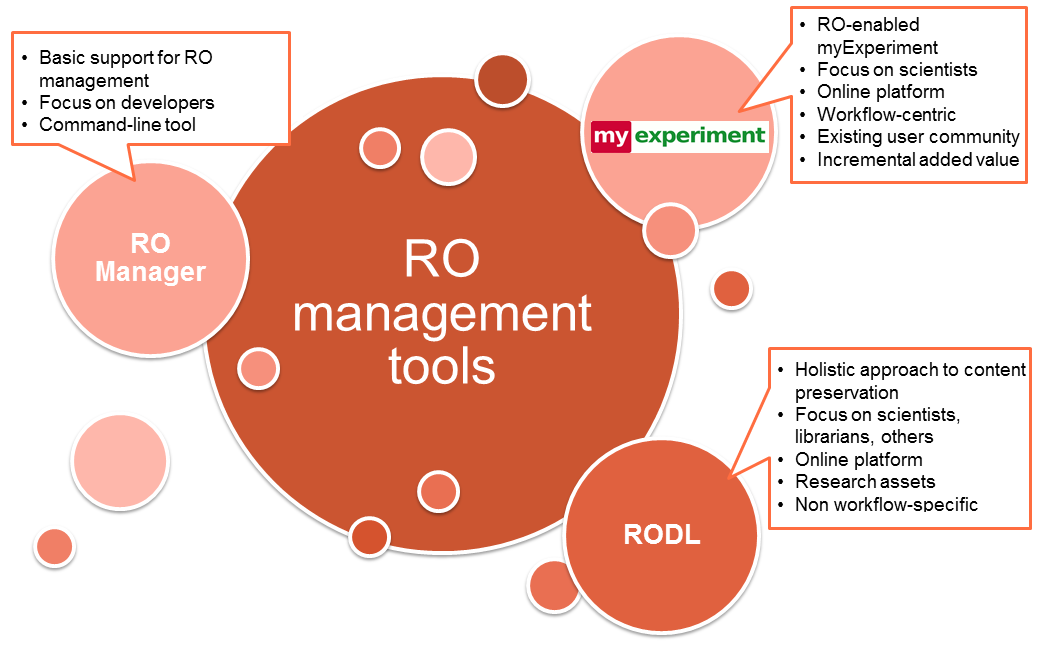}
  \caption{Research Object management tools.}
  \label{fig:romt}
\end{figure*}

To support scientists in creating, annotating, publishing and managing Research Objects, in particular Workflow-Centric Research Objects, we have developed a family of tools. Figure~\ref{fig:romt} illustrates the portfolio of Research Object tools. The tools are aimed towards different types of target users and their needs and represent different levels of deployment of Research Object management capabilities.

The Research Object Manager (RO Manager, described in Section~\ref{subsec:romanager}) is a command line tool for creating, displaying and manipulating Research Objects. The RO Manager incorporates the essential functionalities for Research Object management, especially by developers and a technically skilled audience used to working in a command-line environment. The Research Object Digital Library (RODL, described in Section~\ref{subsec:rodl}) acts as a full-fledged back-end not only for scientists but also for librarians and potentially other communities interested in the aggregation of heterogeneous information sources with the rigor of digital libraries' best practices. RODL provides a holistic approach to the preservation of aggregated information sources and incorporates capabilities to deal  with collaboration, versioning, evolution and quality management of Research Objects. Finally, we have also extended the popular virtual research environment myExperiment \cite{DBLP:journals/fgcs/RoureGS09} to allow end-users who are not necessarily information technology experts, to create, share, publish and curate Workflow-Centric Research Objects (Section~\ref{subsec:myexperiment}). It is worth noting that the developed tools are interoperable. For example, a user can utilise the RO Manager to create Research Objects, and upload them to the RODL portal or myExperiment, where it can undergo further changes.

\subsection{The Research Object Manager}
\label{subsec:romanager}
The Research Object Manager (RO Manager) is a command line tool for creating, displaying and manipulating Research Objects. The RO Manager is complementary to the Research Object Digital Library (RODL) (see Section \ref{subsec:rodl}), in that it is primarily designed to support a user working with Research Objects in the user's local file system. RODL and RO Manager can exchange Research Objects between them, using of the shared Research Object model and vocabularies.  The RO Manager also includes a checklist evaluation functionality, which is used to evaluate if a given Research Object satisfies pre-specified properties (e.g., the input data is declared, the hypothesis of the experiment is present, the Research Object has some examples to play with, etc.).

Experience has shown that a simple command-line tool can provide developers and users with early access to functionality, and provide an opportunity to gather additional user feedback and requirements.  The RO Manager has also been used in conjunction with built-in operating system functionality for scripting prototype tool chains for more complex operations involving Research Objects.

The RO Manager is documented in a user guide that is available online\footnote{\url{http://wf4ever.github.io/ro-manager/doc/RO-manager.html}}.  A FAQ guide describing how to deal with various common operations using RO Manager is also accessible online\footnote{\url{http://www.wf4ever-project.org/wiki/display/docs/RO+Manager+FAQ}}.

The RO Manager is implemented in Python, and it is available as an installable package through the Python Package Index (PyPI)\footnote{\url{https://pypi.python.org/pypi/ro-manager}}. The source code is maintained in the Wf4ever Github repository\footnote{\url{https://github.com/wf4ever/ro-manager}}.

\begin{figure*}[t]
\centering
\includegraphics[width=0.9\textwidth]{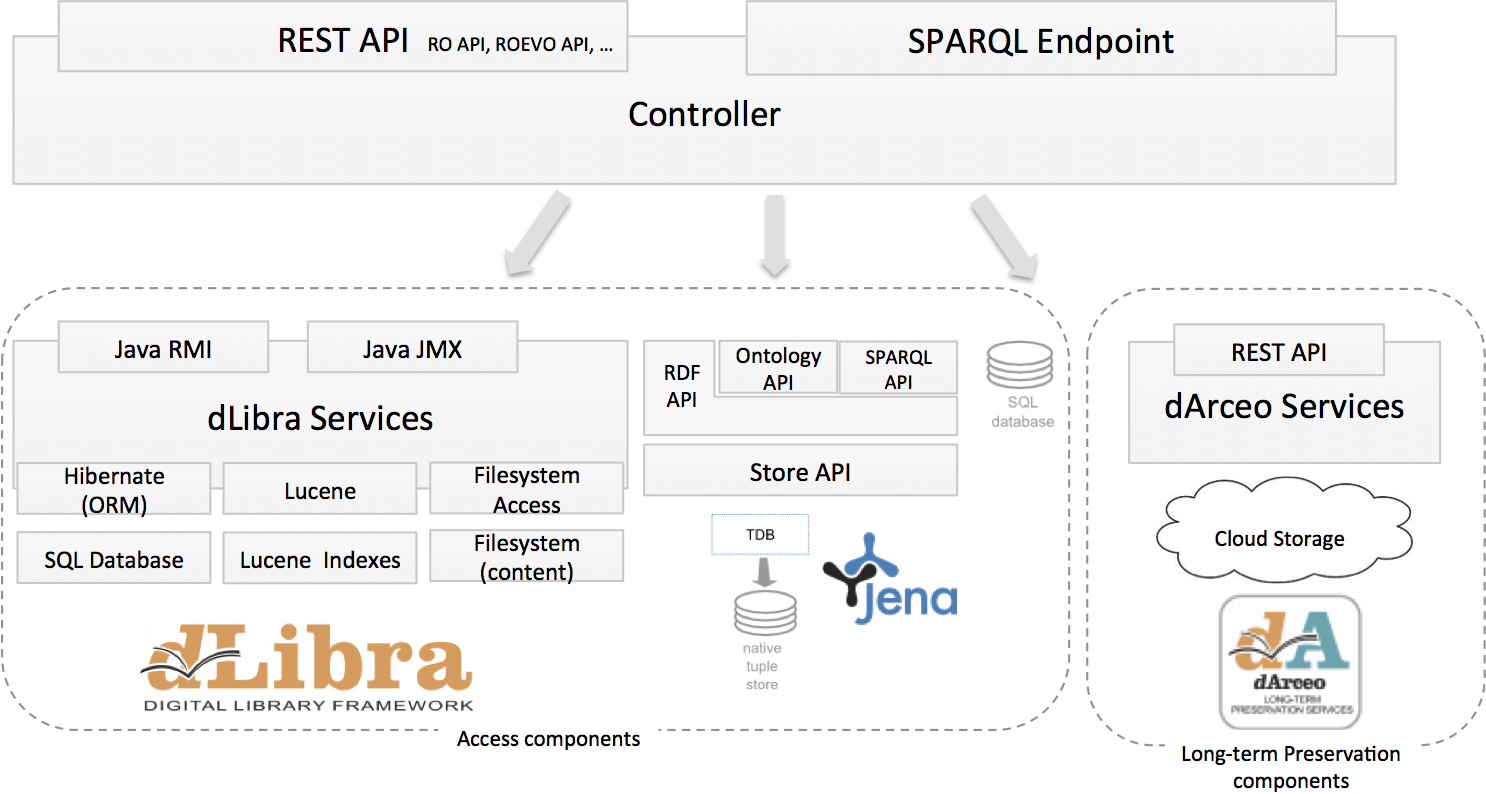}
\caption{Research Objects Digital Library internal component diagram}
\label{RODL}
\end{figure*}

\subsection{Research Object Digital Library}
\label{subsec:rodl}

The foundational service to preserve workflow-centric Research Objects is the Research Object Digital Library (RODL). RODL is a software system which collects, manages and preserves aggregations of scientific workflows and related objects and annotations, packed into Research Objects. RODL is a back-end service that does not directly provide a user interface, but rather system level interfaces through which client software can interact with RODL and provide different user interfaces according to different needs. This section presents the interfaces supported by RODL, describes their implementations and the current clients developed for their Research Object consumption.

\subsubsection{RODL interfaces}

The main system level interface of RODL is a set of REST APIs, including the Research Object  API\footnote{\url{http://wf4ever-project.org/wiki/display/docs/RO+API+6}} and the Research Object Evolution API\footnote{\url{http://wf4ever-project.org/wiki/display/docs/RO+evolution+API}}.

The Research Object API, also called the Research Object Storage and Retrieval API, defines the formats and links used to create and maintain Research Objects in the digital library. It is aligned with the Research Object model that is used to define Research Objects, and so it recognizes concepts such as aggregations, annotations and folders. The Research Object model ontologies are used to specify relations between different resources. Given that the semantic metadata is an important component of a Research Object, RODL supports content negotiation for the metadata resources, including formats such as RDF/XML, Turtle and TriG.

The Research Object Evolution API defines the formats and links used to change the lifecycle stage of a Research Object, most importantly to create an immutable snapshot or archive from a mutable live Research Object, as well as to retrieve the evolution provenance of a Research Object. The API follows the \textit{roevo} ontology (see Section \ref{subsec:roevo}), visible in the evolution metadata generated for each state transition.

Additionally, RODL provides a SPARQL endpoint that allows performing SPARQL queries over HTTP to the metadata of all stored Research Objects\footnote{\url{http://sandbox.wf4ever-project.org/portal/sparql}}. It also implements the Notification API\footnote{\url{http://wf4ever-project.org/wiki/display/docs/Notification+API}}, which defines links used to retrieve Atom feeds with notifications of events about any Research Object. For searching the contents of Research Objects a Solr REST API and the OpenSearch APIs are provided. Finally, RODL implements a custom User Management API\footnote{\url{http://wf4ever-project.org/wiki/display/docs/User+Management+2}} for registering users and generating OAuth 2 access tokens, providing the option of extending it with an access control layer in the future.

\subsubsection{RODL implementation}



One of the main design challenges related to the implementation of RODL was the need to support both live, dynamically changing Research Objects as well as immutable snapshots that are intended for a long-term preservation. As a result,  Figure \ref{RODL} shows an overview of the RODL modular structure, which  comprises the access components (bottom left of Figure \ref{RODL}), the long-term components (bottom right of Figure \ref{RODL}) and the controller that manages the flow of data (top of Figure \ref{RODL}). Immutable Research Objects are stored in the long-term preservation repository once they are created. The live Research Objects, on the other hand, are pushed asynchronously after every change or periodically, depending on the configuration.

The access components are the storage backend - dLibra\footnote{\url{http://dlab.psnc.pl/dlibra/}} - and a triplestore. dLibra provides file storage and retrieval functionalities, including file versioning and consistency checking. It has a built-in text search engine and it manages users and controls their access rights. It allows organizing stored objects into hierarchical structures and associating metadata at the level of object aggregations. It is also possible to use a built-in module for storing Research Objects directly in the filesystem.

The semantic metadata is additionally parsed and stored in the triplestore backed by Jena TDB\footnote{\url{http://jena.apache.org/}}. Jena TDB is an actively developed RDF store implementation, which provides good support for transactions, querying, caching and using named graphs. The use of a triplestore helps in RODL internal data processing and offers a standard query mechanism for RODL clients. It also provides a flexible mechanism for storing metadata about any component of a Research Object identified by a URI, which apart from workflows and other resources, may include parts of workflows or external resources (e.g. web services, data sources).

The long-term preservation component is built on dArceo\footnote{\url{http://dlab.psnc.pl/darceo/}} - a system for long-term preservation of digital objects developed by PSNC. dArceo stores the objects and monitors their quality, alerting the administrators if necessary. The standard monitoring activities include file format decay alerts and fixity checking but can be enhanced using a plugin mechanism. In case of RODL, dArceo periodically monitors the quality of Research Objects by calling the Checklist Evaluation and Stability Services\footnote{\url{http://wf4ever-project.org/wiki/display/docs/RO+checklist+evaluation+API},\url{http://wf4ever-project.org/wiki/display/docs/Stability+Evaluation+API}}. If a change in quality is detected, notifications are generated as Atom feeds in compliance with the Notification API mentioned above. This helps detect and prevent workflow decay which occurs when an external resource or service used by the workflow becomes unavailable or is otherwise behaving differently.


Objects in dArceo can be stored on a range of backends, including specialized preservation repositories such as the Platon service\footnote{\url{http://www.platon.pionier.net.pl/}}, storing data in geographically distributed copies and guaranteeing their consistency.

A running instance of the RODL is available for testing\footnote{\url{http://sandbox.wf4ever-project.org/rodl/}}. At the moment of writing, it holds more than 1100 Research Objects.

\begin{figure*}[t]
\begin{center}
\includegraphics[width=0.8\textwidth]{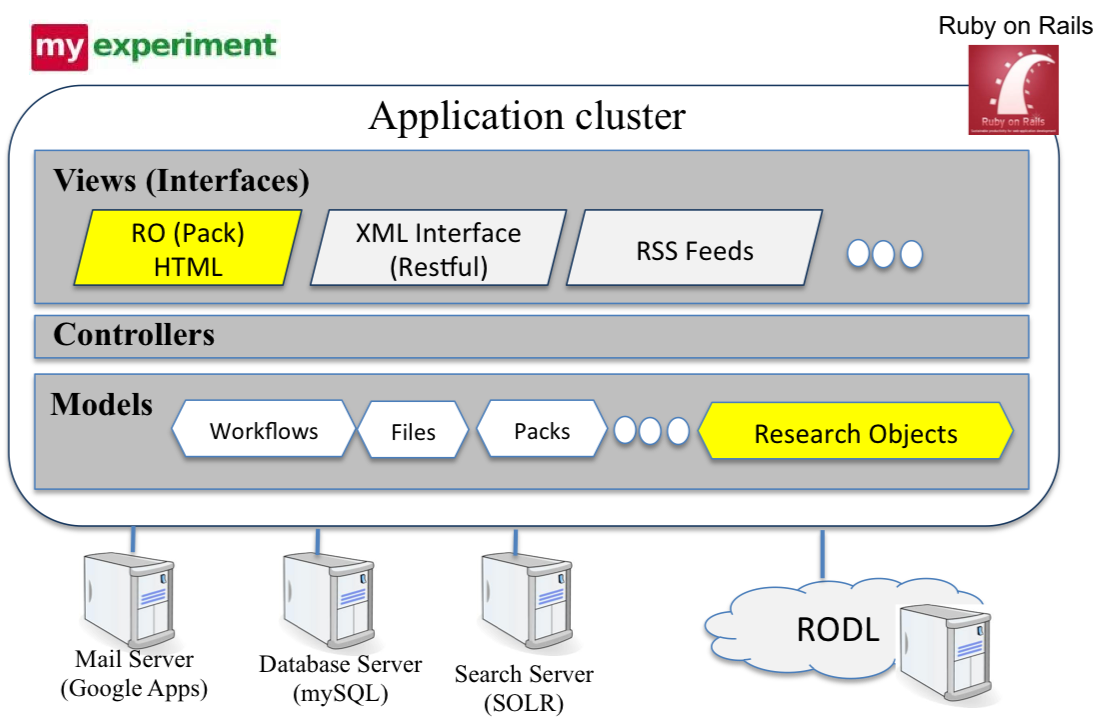}
\end{center}
\caption{RO-enabled myExperiment.}
\label{fig:myexperimentarchitecture}
\end{figure*}

\subsubsection{RODL clients}

The reference client of RODL is \textbf{the Research Object Portal (RO Portal)}, developed alongside RODL to test new features and expose all available functionalities. It is running as a web application\footnote{\url{http://sandbox.wf4ever-project.org/portal}}. Its main features are Research Object exploration and visualization, but it also allows creating user accounts in RODL and generating access tokens for other clients. The RO Portal uses all APIs of RODL. 
The development version of \textbf{myExperiment} (see Section \ref{subsec:myexperiment}) 
 uses RODL as a backend for storing packs. It uses the Research Object API. Finally, the \textbf{RO Manager} (see Section \ref{subsec:romanager}) 
 allows to push a Research Object to RODL via the Research Object API, as well as converting it into a snapshot in RODL.

\subsection{Research Object-Enabled myExperiment}
\label{subsec:myexperiment}


myExperiment \cite{DBLP:journals/fgcs/RoureGS09} is a virtual research environment targeted towards collaborations for sharing and publishing workflows (and experiments). It provides the functionalities necessary for sharing workflows within and across multiple communities. In doing so, myExperiment adopts a social web approach, which is adapted to the need of scientists. The workflows that are shared using myExperiment do not need to be specified in a particular workflow management system. For example, we find on myExperiment workflows that have been specified using Galaxy \cite{galaxy}, Taverna \cite{taverna}, Kepler \cite{kepler} and Vistrails \cite{vistrails}.

While initially targeted towards workflows, the creators of myExperiment were aware that scientists want to share more than just workflows and experiments. Because of this, myExperiemnt was extended to support the sharing of artifacts (known as Packs). A pack can be seen as a basic aggregation of resources, which can be workflows, files, presentations, papers, or links to external resources. 
The notion of packs has been widely adopted by scientists. At the time of writing, myExperiment had $337$ packs. Just like a workflow, a pack can be annotated and shared.

In order to support complex forms of sharing, reuse and preservation, we have incorporated the notion of Research Objects (which can be seen as advanced packs) into the development version of myExperiment \footnote{\url{http://alpha.myexperiment.org/packs/}}. In addition to the basic aggregation supported by packs, alpha myExperiment provides the mechanisms for specifying metadata that describes the relationships between the resources within the aggregation. Moreover, the structure and the types of the resources that compose a pack are now inline with those that have been identified thanks to the Research Object model. For example, a user is able to specify that a given file within a pack is the hypothesis, that another file specifies the workflow run obtained by enacting a given workflow, or that a given file states the conclusions drawn by the scientists after analyzing the workflow run.


Figure \ref{fig:myexperimentarchitecture} illustrates a high-level architecture of alpha myExperiment. As illustrated in the figure, at the level of the Rails\footnote{\url{http://rubyonrails.org}} model, data structures that represent the Research Object and associated resources have been incorporated. To manipulate such data structures, the controller layer has been extended, and to provide non-information technology users with the ability to create and manage Research Objects, the view layer has been extended with the necessary HTML Web pages.

To illustrate how myExperiment can be used for managing Research Objects, consider that the user (Alice) wants to create and share a Research Object. To do so, Alice first browses myExperiment to identify a workflow that is of interest to her investigation. Once she identifies a relevant workflow, she downloads the workflow, modifies and re-purposes it for her investigation. Once she is happy with the \emph{new} workflow, Alice decides to create a Research Object. In doing so, she specifies the hypothesis within a file, which is stored within RODL. RODL acts as a back-end for myExperiment to store the information about Research Objects. Alice then uploads her workflow to myExperiment. As a result, myExperiment sends a request to the \texttt{Research Object transformation service}, which uploads the workflow definition to RODL, transforms the workflow definition according to the wfdesc ontology, and extracts the annotations that are bundled within the workflow definition. These elements, i.e., wfdesc specification and annotations, are then uploaded to the Research Object in RODL. Alice also uploads the workflow runs obtained as a result of enacting her workflow, and specifies the conclusion she comes to at the end of her investigation. 

Using myExperiment, Alice now has a Research Object, compliant with the models presented in Section \ref{sec:ontologies_overview}, viewable and manipulable as a pack through myExperiment, and enriched with a hypothesis and conclusions that can assist other users in understanding and possibly reusing and re-purposing her research results.

%% file: conclusions.tex
\section{Conclusions} 
\label{sec:conclusions}

Scholarly articles are in many cases not enough for sharing complete investigation results. In this paper, we presented a family of ontologies that realize the Research Object vision set by Bechhoffer {\em et al.} \cite{DBLP:journals/fgcs/BechhoferBRMABCCDDGMONSG13} to enable the interpretation, reuse and reproducibility of the results of scientific investigations. The Research Object ontologies are organized into a core vocabulary, for annotating and aggregating research investigation resources into meaningful bundles, and extensions vocabularies, for specifying workflows (experiments), the provenance of their executions and the evolution of the Research Object over time. Furthermore we reported on available tools that can be used by scientists to effectively create, share, and preserve their Research Objects through repositories like myExperiment.

While the notion of Research Object was initially developed under the aegis of the Wf4Ever project\footnote{\url{http://www.wf4ever-project.org}}, its ethos, models and tools are being adopted and exploited by other communities. In particular, the BioVel\footnote{\url{http://www.biovel.eu}} and Scape\footnote{\url{http://www.scape-project.eu}} projects are using Research Objects to support the annotation and preservation of workflow specifications together with their workflow runs.  In our ongoing work, we seek to collaborate with these communities, as well as others who we are working with, e.g., Timbus\footnote{\url{http://timbusproject.net}} and GigaScience\footnote{\url{http://www.gigasciencejournal.com}}, to improve the Research Object vocabularies and tools in the light of the feedback and lessons we learn through interactions with these communities. 

Finally, since Research Objects have demonstrated to be valuable resources for many different communities when sharing scholarly publications, we proposed to start a W3C community group\footnote{\url{http://www.w3.org/community/rosc}} to gather additional use cases and discussions that would improve the current model. The group has been approved and more than 80 participants from different organizations have signed in as contributors\footnote{\url{http://www.w3.org/community/rosc/participants}}.